# Deploying large fixed file datasets with SquashFS and Singularity

Pierre Rioux[1,†]
pierre.rioux@mcgill.ca

Gregory Kiar[1]
gkiar@mcin.ca

Alexandre Hutton[1]
ahutton@cim.mcgill.ca

Alan C. Evans[1]
alan@bic.mni.mcgill.ca

Shawn T. Brown[1,2]
stbrown@psc.edu

[1]Montreal Neurological Institute, McGill University, Montreal, QC, Canada
[3]Pittsburgh Supercomputing Center, Pittsburgh, PA, USA

## ABSTRACT

Shared high-performance computing (HPC) platforms, such as those provided by XSEDE and Compute Canada, enable researchers to carry out large-scale computational experiments at a fraction of the cost of the cloud. Most systems require the use of distributed filesystems (e.g. Lustre) for providing a highly multi-user, large capacity storage environment. These suffer performance penalties as the number of files increases due to network contention and metadata performance. We demonstrate how a combination of two technologies, Singularity and SquashFS, can help developers, integrators, architects, and scientists deploy large datasets (O(10M) files) on these shared systems with minimal performance limitations. The proposed integration enables more efficient access and indexing than normal file-based dataset installations, while providing transparent file access to users and processes. Furthermore, the approach does not require administrative privileges on the target system. While the examples studied here have been taken from the field of neuroimaging, the technologies adopted are not specific to that field. Currently, this solution is limited to read-only datasets. We propose the adoption of this technology for the consumption and dissemination of community datasets across shared computing resources.

## 1 Introduction

As large-scale computing and data resources have become increasingly accessible, many domains of science have shifted their focus towards the collection, pooling, dissemination, and shared consumption of large data resources. With the Sloan Digital Sky Survey [14], Open Connectome Project [3], Human Genome Project [4], Allen Human and Mouse Brain Atlases [5, 11], many domains of science have benefited from grass-roots efforts at providing rich and expansive datasets for their shared exploration. While in many cases these initiatives provide federated access mediated by access-controlled databases, it often remains impractical to connect these repositories directly to computing infrastructure (e.g. when computing nodes restrict internet access).

This limitation necessitates the local cloning of large shared resources. While these may come in the form of small collections of large files, such as is often the case for high resolution imaging data [3], cohort studies involving many individuals, such as the Human Connectome Project (1,200 participants) [7] or UK BioBank (500,000 participants) [13] often consist of much larger collections of considerably smaller files. Though application developers familiarize themselves with the data organization that concerns them and create their tools accordingly, these data organizations are not equivalent from the perspective of HPC systems.

While obtaining storage devices that have the capacity to store thousands of TB of data has become relatively inexpensive, the organization of these devices and structure of the information being stored has large implications on the system's performance. Problems start to occur when several million files are present: despite the filesystem's capability for successfully managing and storing them all, access can become prohibitively slow. User processes (like scripts, scientific pipelines, or even system backup utilities) often scan the list of files in a sequential manner. This can happen even if the processes don't intend to open all the files for reading: the file list (their names and attributes) is a first step in deciding what to do later on. While there are programmatic methods for improving the I/O performance at an application level (parallelism in file reading and writing), it is still a challenge to provide a general solution for many common operations that does not require rewriting software.

At 10,000 files per second, a reasonable speed demonstrated in following experiments, it would take 10 minutes to scan 6 million files. If a dataset contains, say, 100 million files, then a process scanning them will take nearly 3 hours to go through the list without any additional processing. Looking at the multi-modal UK BioBank dataset containing 500,000 subjects, each with approximately 60 files of data, this raw collection initially consists of 30 million files which explodes to over 30 billion once the data is processed using standard tools [6], not including perturbations that may be carried out in parallel by different researchers. This derivative dataset would take weeks to scan, rendering any realistic attempt at performing backups or processing futile.

The problem stated above is accentuated when the files are stored on shared, networked filesystems, known as distributed filesystems. In this model, a large pool of file storage is configured on high speed, high reliability dedicated storage systems, and exported via NFS [10], Lustre [2], or GlusterFS [1], to a set of computation nodes. Most scientific computing clusters use such distributed filesystems. The nodes are shared among many users who run different applications on them. The distributed filesystem has to reply and answer in parallel to the file access requests of all these applications, involving locking of metadata information about the files so that the structure can be consistently maintained for all users [12]. Even with sizable caching at all levels of the architecture, it is often not possible to avoid significant slowdown in replying to the requested file operations for all those applications.

The method presented here addresses this issue through the creation of read-only filesystem images and the overlay of these images in virtual environments. While the technological components used in this work are not novelly claimed here, the authors believe that this combination of technologies is novel and uniquely enables large scale scientific exploration on shared computing resources with minimal burden on administrators, users, and the distributed filesystem infrastructures themselves.

## 2   Materials & Methods

The core contribution of this paper is a novel application of existing technologies to enable the scalable consumption and dissemination of datasets on shared resources. This was accomplished through two core tools which will be introduced in the following section: SquashFS [9], and Singularity [8]. All wrappers and scripts produced as a part of this work can be found publicly available on GitHub at https://github.com/aces/sing-squashfs-support.

### 2.1   SquashFS

SquashFS [9] is a POSIX-compliant filesystem format designed to pack read-only files and directories into a single contiguous segment of bytes. It is a stable image format and has built-in support in the Linux kernel. It was designed to efficiently store system files for live CDs, small device firmwares, and other embedded systems. It can also be used to store arbitrary datasets as long as the files are not expected to change.

The mksquashfs utility builds a SquashFS filesystem. It takes an arbitrary set of files and folders and packs them into a single large normal file: a SquashFS filesystem file. Accessing the packed files afterwards is performed by creating a mount point and asking the kernel to mount the SquashFS file. Several packing options are supported, such as compression of data block or inodes, but they are transparently recognized when the filesystem is mounted. The action of mounting normally requires privileged (root) access, meaning while a standard user can build a SquashFS filesystem they cannot expect to be able to mount it.

### 2.2   Singularity

Singularity [8] is a container technology for Linux. Given an image of a Linux installation (system files, application code), it can start user processes in a Linux container where the filesystems seen by these processes are limited to those in the image. Although the Singularity launcher itself is installed with root privileges, a non-privileged user can start a container without requiring (or gaining) root privileges; processes will always inherit the user's privileges.

Alongside those belonging to the Singularity image itself, users are able to specify additional filesystems to bring into the container at start time. These appear as new directories in the filesystem namespace; running contained processes can access them as perfectly traditional files and folders. As explained above, normally on Linux, mounting a filesystem is a procedure that can only be performed by the root user. This feature of Singularity means that a non-privileged user can launch a process that will have access to externally-mounted filesystems, as long as that process is containerized, even though they are not mounted on the host computer. This side effect is convenient if the user happens to be launching an application on a cluster where the user does not normally have root access.

Singularity calls its filesystem mounting feature overlays. Overlays are provided as filesystems-within-a-file. A normal file is prepared in advance, packing a filesystems's set of files. When a Singularity container starts it will mount the filesystem internally. As of 2020Q1, Singularity supports ext3 and SquashFS filesystem overlays.

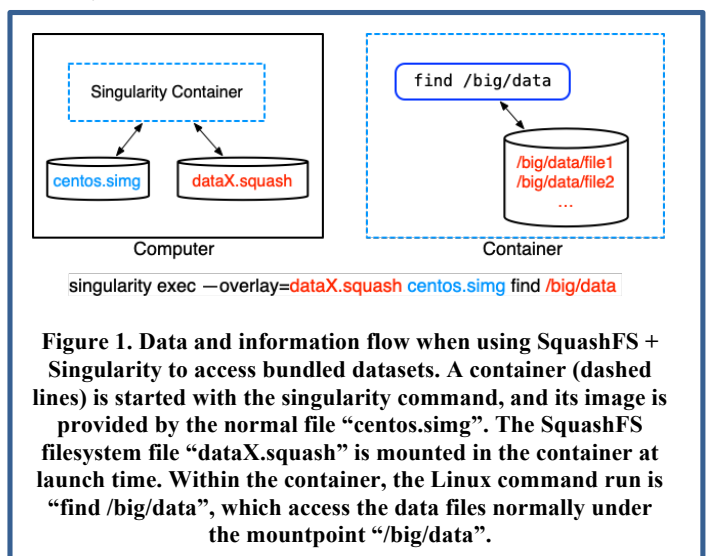

**Figure 1. Data and information flow when using SquashFS + Singularity to access bundled datasets. A container (dashed lines) is started with the singularity command, and its image is provided by the normal file "centos.simg". The SquashFS filesystem file "dataX.squash" is mounted in the container at launch time. Within the container, the Linux command run is "find /big/data", which access the data files normally under the mountpoint "/big/data".**

### 2.3   Combining Singularity & SquashFS

SquashFS filesystems accessed through Singularity provides the means for a non-privileged user to mount a packed filesystem in

Deploying large fixed file datasets with SquashFS and Singularity

Linux . Our GitHub repository contains flexible wrappers which enable this workflow in a variety of settings. They provide automatic detection of the SquashFS files and an associated singularity image and built-in support for transparent file access, sshfs, SFTP, rsync, and other interactive and non-interactive commands.

In the simplest use case, a scientific application and dataset are already fully packaged in a Singularity image and SquashFS image, respectively. In this case, the Singularity image only needs to be launched with the appropriate overlay to access the packed files of the SquashFS filesystem. Figure 1 shows how information is routed using this method.

Figure 2A shows a traditional mounting process for sshfs, allowing users to access remote datasets. Figure 2B shows the prerequisites for performing a sshfs mount using a SquashFS filesystem and Singularity. Finally, Figure 2C adapts Figure 2A to illustrate how a user can access a dataset inside SquashFS files transparently through system tools, both on the host and inside the container, and mount the dataset remotely as though it were a typical volume.

The wrappers illustrated above enable the efficient and transparent access of large datasets on both local and shared computing resources. In the following section we demonstrate the efficacy of this approach compared to a naive access strategy.

## 3 Results

The method presented above was tested using a large packaged dataset on one of the supercomputer clusters of Compute Canada (a federal organization providing several HPC environments to the Canadian academic community). Compute Canada's platforms have strict disk quotas that limit the number of files users can place on them, similar to many shared systems, restricting individual research groups to several million files to avoid many of the concerns raised above.

The Human Connectome Project (HCP) dataset [7] has file-level summary statistics shown in Table 1. The dataset was packed into 56 SquashFS files, each containing up to 20 of the total 1113 subjects, averaging 1.5 terabytes each. While these bundles still accounted for 88 TB of space on disk they were contained within approximately 1 file per 300,000 in the original collection. Alongside those files, we installed a README.txt and a set of utility wrappers to help users access the data files.

**Table 1. Storage properties of the HCP dataset.**

| Raw Dataset | |
|---|---|
| Number of files | 15,716,005 |
| Number of directories | 940,082 |
| Depth of directory structure | 7 |
| Total size | 88,577,644,617,358 bytes (88.6 TB) |
| | |
| Bundled Dataset | |
| Number of SquashFS files | 56 |
| Total size | 87,171,340,062,720 bytes (87.2 TB) |
| Average file size | 1.5 TB |

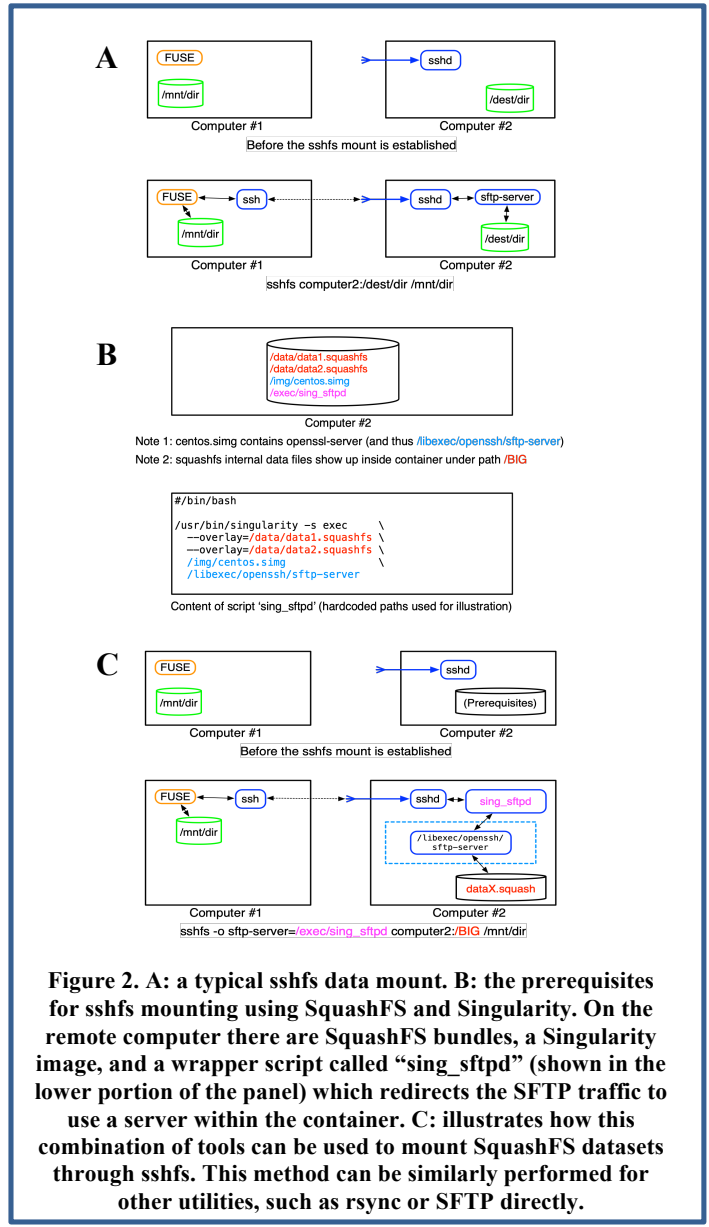

Figure 2. A: a typical sshfs data mount. B: the prerequisites for sshfs mounting using SquashFS and Singularity. On the remote computer there are SquashFS bundles, a Singularity image, and a wrapper script called "sing_sftpd" (shown in the lower portion of the panel) which redirects the SFTP traffic to use a server within the container. C: illustrates how this combination of tools can be used to mount SquashFS datasets through sshfs. This method can be similarly performed for other utilities, such as rsync or SFTP directly.

### 3.1 Boot Performance

Starting a Singularity container with a local Singularity image with no overlays typically takes on the order of a second. When adding SquashFS files as overlays the boot time increases based on both the size of the overlay and how many are loaded. For the 1.5 TB SquashFS overlays described above the delay can be as large as about 1 second per overlay. This delay is lessened when the data blocks of the SquashFS files are already cached by the host. In the case of the 88 terabytes HCP dataset, the container becomes available after approximately a 1 minute delay during a fresh boot session but less than 2 seconds when re-launched again immediately.

## 3.2 File Access Performance

Table 2 shows performance tests made with the HCP dataset. The basic test consists of listing all the files and directories and counting them. The tests were performed twice in a row to properly highlight the effects of caching by the host computer. The test was launched as 42 cluster jobs over a period of two days, which ended up being distributed to 7 different computing nodes. A cluster job included three such pairs of tests, on:

1. a subset of the HCP dataset as normal files on the filesystem of the cluster, containing 186,432 files and directories (1% of the full dataset),
2. the same subset of HCP but stored on SquashFS and accessed with Singularity, and
3. the full HCP 1200 dataset with SquashFS and Singularity

The min and max times of each collection of times was removed, and the average of the 40 remaining time values are shown in Table 2. The values in this table show that file access performance increases by 6–10x through the use of SquashFS + Singularity in both the first and second scan settings. This performance increase is maintained not only for a dataset of the equivalent size, but scales to the entire HCP collection that is 100x larger than the collection tested on the host system.

**Table 2. Scan time performance to inspect the directory tree in three different environments using the Linux command "time (find . –print | wc –l)", averaged across 40 runs.**

|        | 1% HCP Subset |          | Full HCP |
|--------|---------------|----------|----------|
|        |               | SquashFS |          |
| Files  | 186,432       | 186,432  | 16,656,087 |
| Scan 1 | 12.9 seconds  | 2.1 seconds | 147.4 seconds |
|        | 14.5K entries/s | 88.4K entries/s | 113.0K entires/s |
| Scan 2 | 5.0 seconds   | 0.6 seconds | 66.9 seconds |
|        | 37.2K entries/s | 309.3K entries/s | 248.8K entries/s |

## 4 Discussions

We believe that the joint application of SquashFS and Singularity presented here provide users, administrators, and dataset creators with a scalable method for creating, disseminating, and consuming large collections of file-based data. Within the container, file access is fast both in with respect to the content of the files and the meta information (inode data) obtained from Linux system calls such as readdir() and stat(). This is a great advantage of the packing solution described in this paper. All file access primitives made inside the container are essentially transformed into sequences of llseek() and read() system calls made by the host kernel on the SquashFS files content. Given that the host's kernel will aggressively cache that information, and that all the inode information is localized within the SquashFS files, it means the basic information about the dataset files become quickly cached even with millions of files.

One limitation of the approach presented here is the dependence on read-only SquashFS bundles. While Singularity also supports overlay files formatted in ext3, which are writable, it is important to note that their size must be pre-allocated, limiting their usefulness. Ext3 filesystems can be created and mounted without requiring root access, and mounted separately or on top of SquashFS filesystems. A possible application of an ext3 system on top of a SquashFS overlay would be to allow the modification of original data such that the versions on the ext3 system supersede the original. We are planning to incorporate such features in the wrapper scripts provided to the community. However, a limitation of ext3 filesystems are that at most one Singularity container may mount them at any given time, unlike for SquashFS. Currently, the authors propose that derived datasets be stored on typical host filesystems rather than anticipating and pre-allocating storage in an ext3 filesystem; these files could then be bundled and stored in an ext3 system post-hoc, once storage requirements are definitively known.

SquashFS is not normally considered as a technology for storing large sets of files, but in combination with Singularity it becomes an effective solution for packing datasets and making them available to user processes. Importantly, this approach does not require elevated privileges and is accessible by standard users. The adoption of bundle-based dataset administration presented here has the potential to greatly increase the performance of shared distributed filesystems with minimal overhead for users and administrators.

## ACKNOWLEDGMENTS

The author would like to thank the people who provided feedback on this paper, namely Natacha Beck and the developers working on Singularity, who fixed a regression bug in versions 3.5.[0-2] while this paper was being written. Funding for this work was provided by CFREF/HBHL (Canada First Research Excellence Fund/Healthy Brains for Healthy Lives). Data were provided in part by the Human Connectome Project, WU-Minn Consortium (Principal Investigators: David Van Essen and Kamil Ugurbil; 1U54MH091657) funded by the 16 NIH Institutes and Centers that support the NIH Blueprint for Neuroscience Research; and by the McDonnell Center for Systems Neuroscience at Washington University.

Deploying large fixed file datasets with SquashFS and Singularity